\documentclass[12pt]{iopart}
\usepackage{amssymb,graphicx,amsbsy}

\begin{document}

\title[Hierarchical renormalization group on bond percolation]{Hierarchical
renormalization-group study on the planar bond-percolation problem}
\author{Seung Ki Baek and Petter Minnhagen}
\ead{garuda@tp.umu.se}
\address{Integrated Science Laboratory, Department of Physics, Ume{\aa}
University, 901 87 Ume{\aa}, Sweden}

\begin{abstract}
For certain hierarchical structures, one can study the percolation problem
using the renormalization-group method in a very precise way.
We show that the idea can be also
applied to two-dimensional planar lattices by regarding them as hierarchical
structures. Either a lower bound or an exact critical probability can be
obtained with this method and the correlation-length critical exponent is
approximately estimated as $\nu \approx 1$.
\end{abstract}


\maketitle

The percolation problem is a question about how a global connection
can be made possible by randomly filling local components by a certain
probability $p$. While it can be explained in purely geometric terms without
any interaction, when a global connection actually appears, the macroscopic
behavior of the system exhibits all the characteristic features of a continuous
phase transition with a diverging correlation length, just as we observe in
other interacting spin systems such as the two-dimensional (2D) Ising
model~\cite{stauffer}.
This analogy is given a precise meaning by the Fortuin-Kasteleyn
representation of the $q$-state Potts model~\cite{fk}, where the percolation
turns out to be equivalent to the limit of $q \rightarrow 1$. Since the
percolation transition at a critical probability $p_c$ has a diverging
correlation length, every microscopic length scale becomes irrelevant with
respect to the critical phenomena, and the system behaves as if it does not
have any specific length scale. This is a qualitative explanation of the
reason why a percolating cluster connecting two opposite sides of a 2D plane
has a fractal dimension at $p=p_c$. The lack of a specific length scale
implies that the system remains statistically invariant even if we zoom the
system up or down, and this scale invariance readily lends itself to a
renormalization-group (RG) study of the percolation
problem~\cite{nie,rey77,rey80}.

\begin{figure}
\begin{center}
\includegraphics[width=0.35\textwidth]{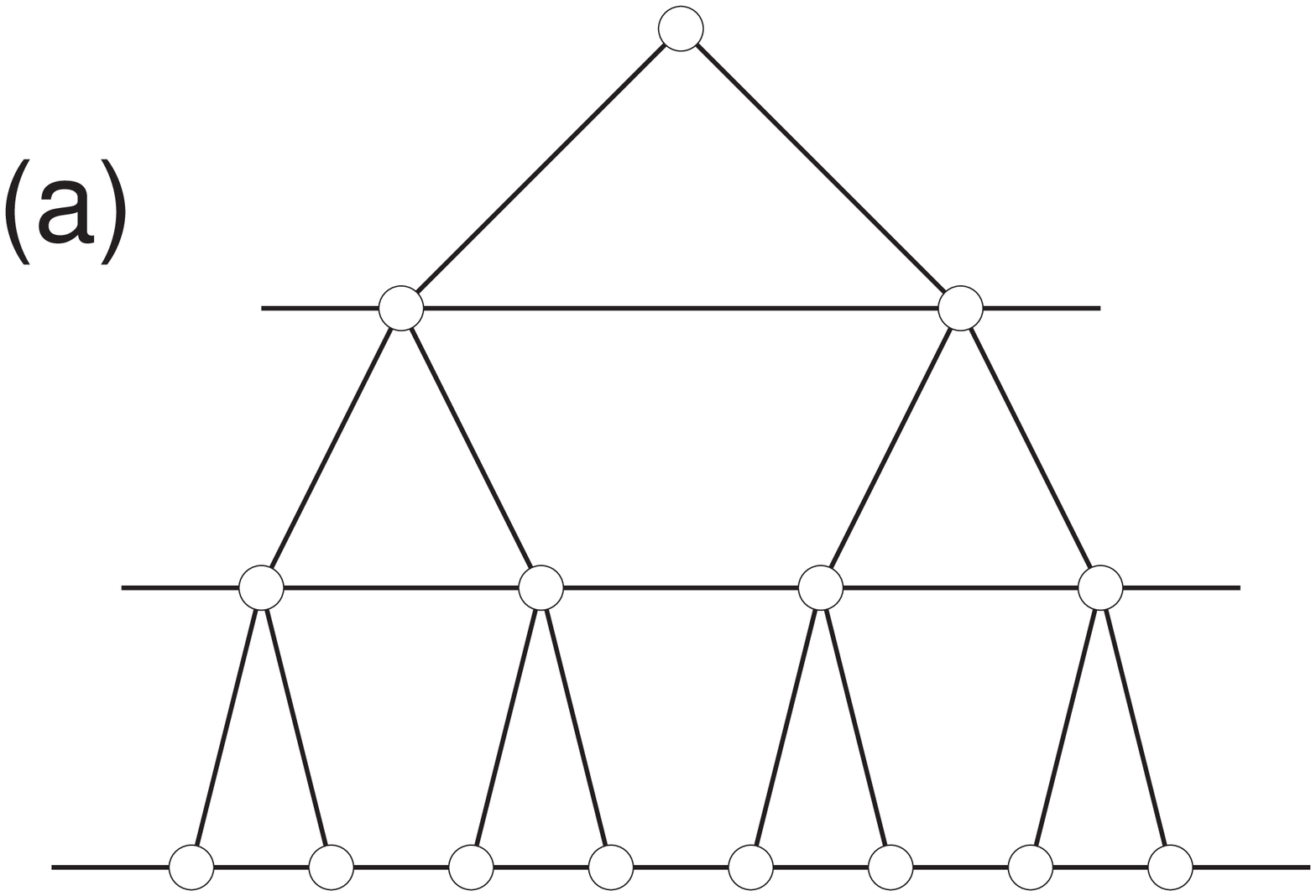}
\includegraphics[width=0.28\textwidth]{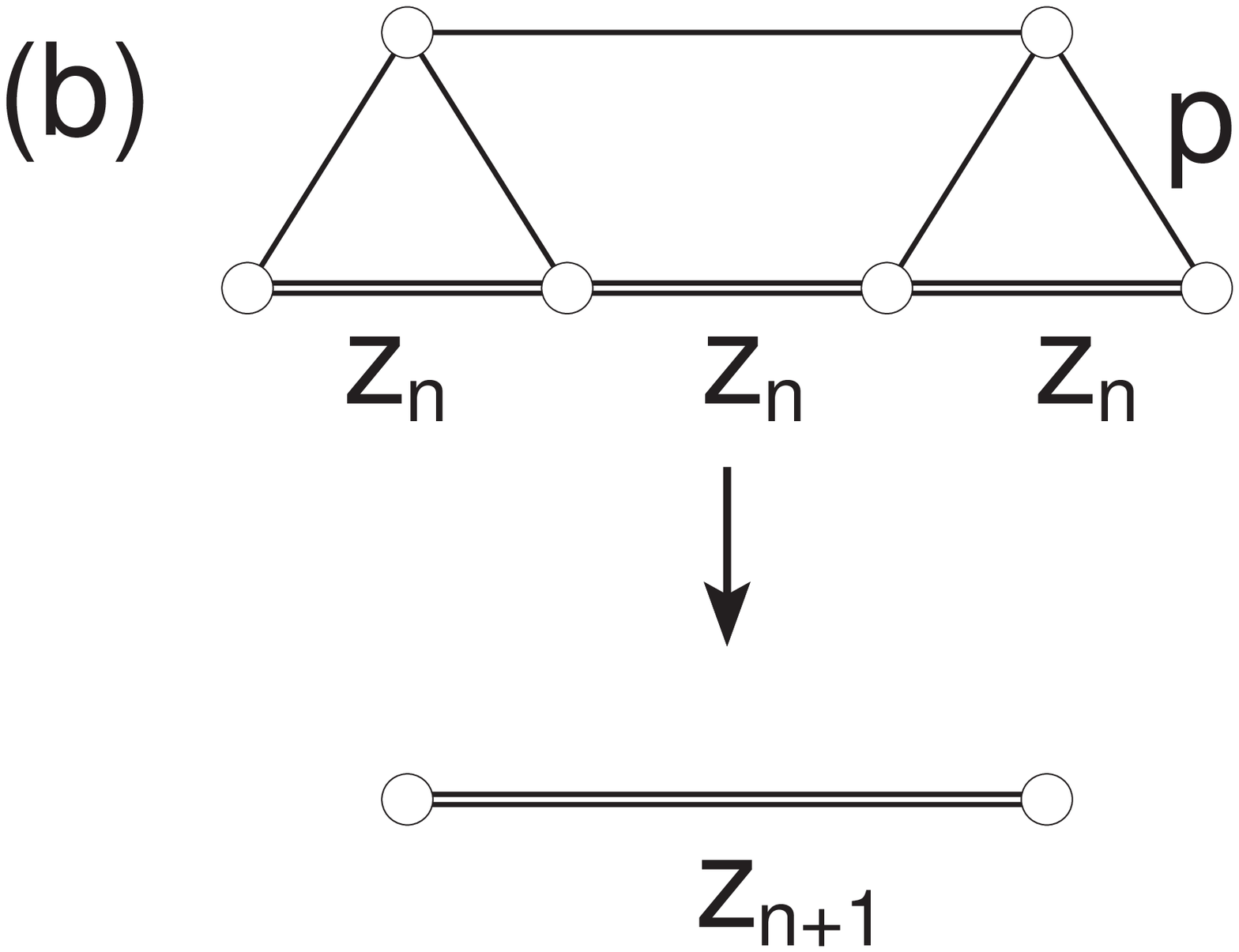}\\
\includegraphics[width=0.41\textwidth]{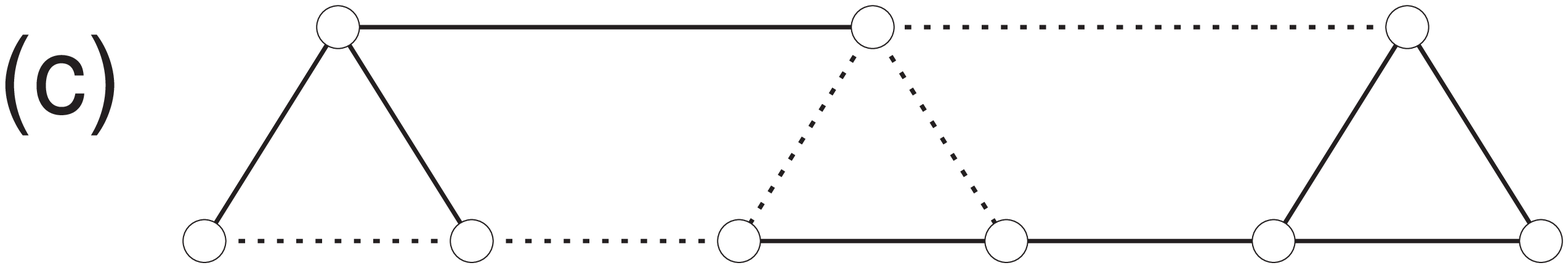}
\end{center}
\caption{
(a) Enhanced binary tree, a hierarchical structure derived from a simple
tree with branching number 2.
(b) RG scheme for the enhanced binary tree, where the
connectivity over the cell above is coarse-grained into a single bond filled
by probability $z_{n+1}$.
(c) This layer is not connected from left to right even though the
cells inside it appear as filled according to the recursion scheme in (b).
The solid and dotted lines represent filled and empty bonds,
respectively.
}
\label{fig:ebt}
\end{figure}

In certain cases where the underlying structure itself is fractal, it is
possible to carry out the RG calculation to a good
approximation or exactly, exploiting this fractal
property~\cite{rozen,boet}.
Such fractal structures usually contain groups of bonds which connect longer
and longer distances in a regular fashion. For this reason, one can
sometimes arrange the groups of bonds in a hierarchical way according to their
connection lengths. \Fref{fig:ebt}(a) is an example of a hierarchical
structure called the enhanced binary tree, which is obtained by adding
horizontal bonds to the simple binary tree. It is hierarchical in the sense
that filling a horizontal bond is comparable to a very long connection along
the bottom layer and the connection length is dependent on the level of the
horizontal bond~\cite{ebt}. That is, a horizontal bond in the highest level
can connect two points at distance 7 along the bottom layer at maximum. For
a horizontal bond at the next highest level, this maximum connection
distance is only as large as 3 lattice spacings. An
RG scheme for the enhanced binary tree is described in \cite{ebt} as
shown in \fref{fig:ebt}(b): we calculate the probability for any of the
leftmost points to connect to any of the rightmost points within the cell
as a function of the bare coupling $p$ and a coarse-grained effective
coupling $z_n$, and then replace this probability by a new effective bond
with strength $z_{n+1}$. The resulting expression for $z_{n+1}$ is written
as
\begin{equation*}
z_{n+1} = p + (1-p) \left[ (1-p)^2 z_n^3 + 2p(1-p) z_n^2 +
p^2 z_n \right].
\end{equation*}
By asking when $z_n = z_{n+1} = z_{\infty}$ becomes 1, we
obtained a lower bound of the percolation threshold as $p_c \ge
1/2$~\cite{ebt}, which is consistent with the conclusion in \cite{future}
that $p_c=1/2$.
Note that we get a lower bound since in iterating $z_n$ to
$z_{n+1}$, there is a small chance to regard a layer as percolated when it
is actually not [see, e.g., \fref{fig:ebt}(c)], whereas the opposite is
not possible.

\begin{figure}
\begin{center}
\includegraphics[width=0.28\textwidth]{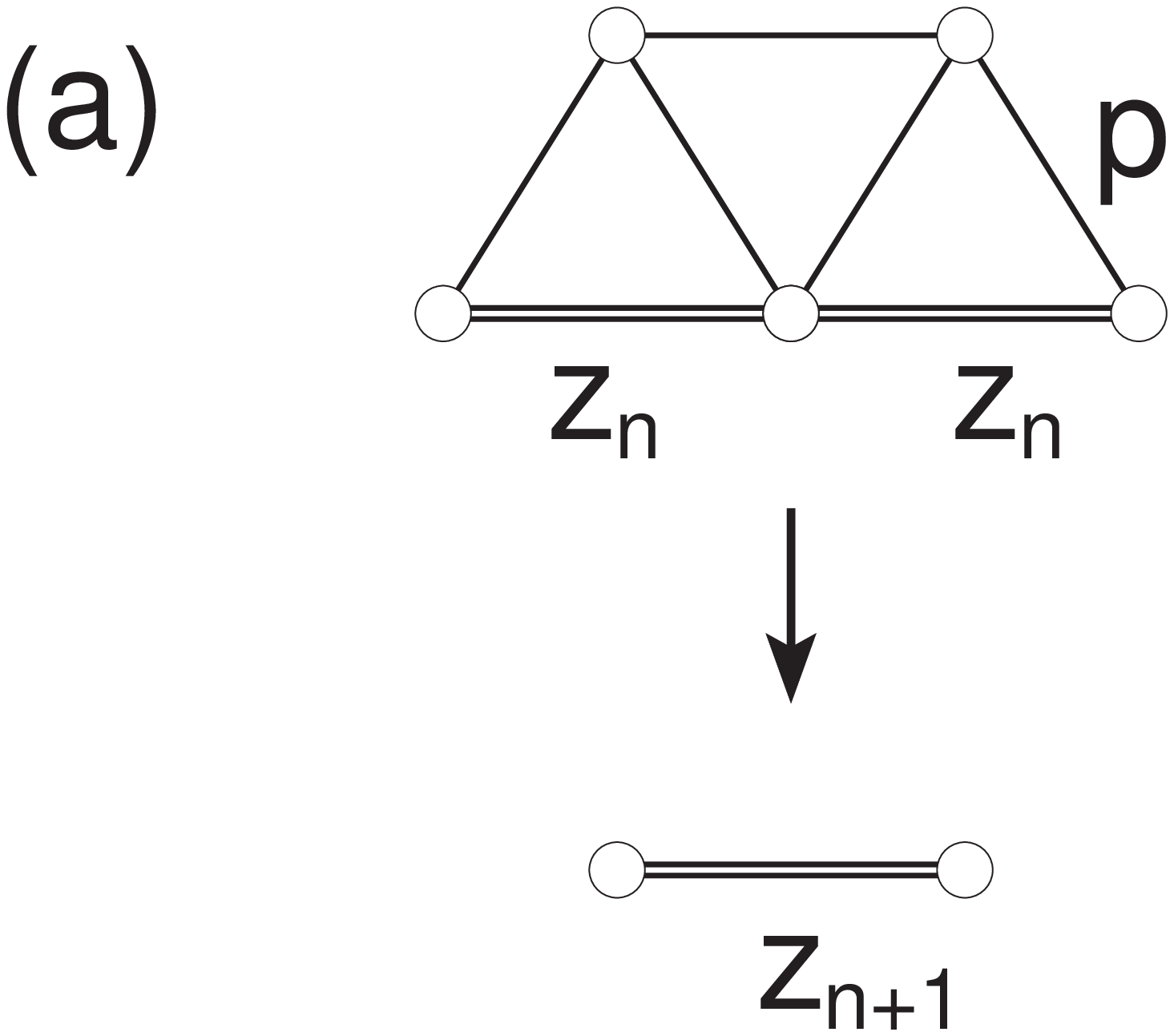}
\includegraphics[width=0.28\textwidth]{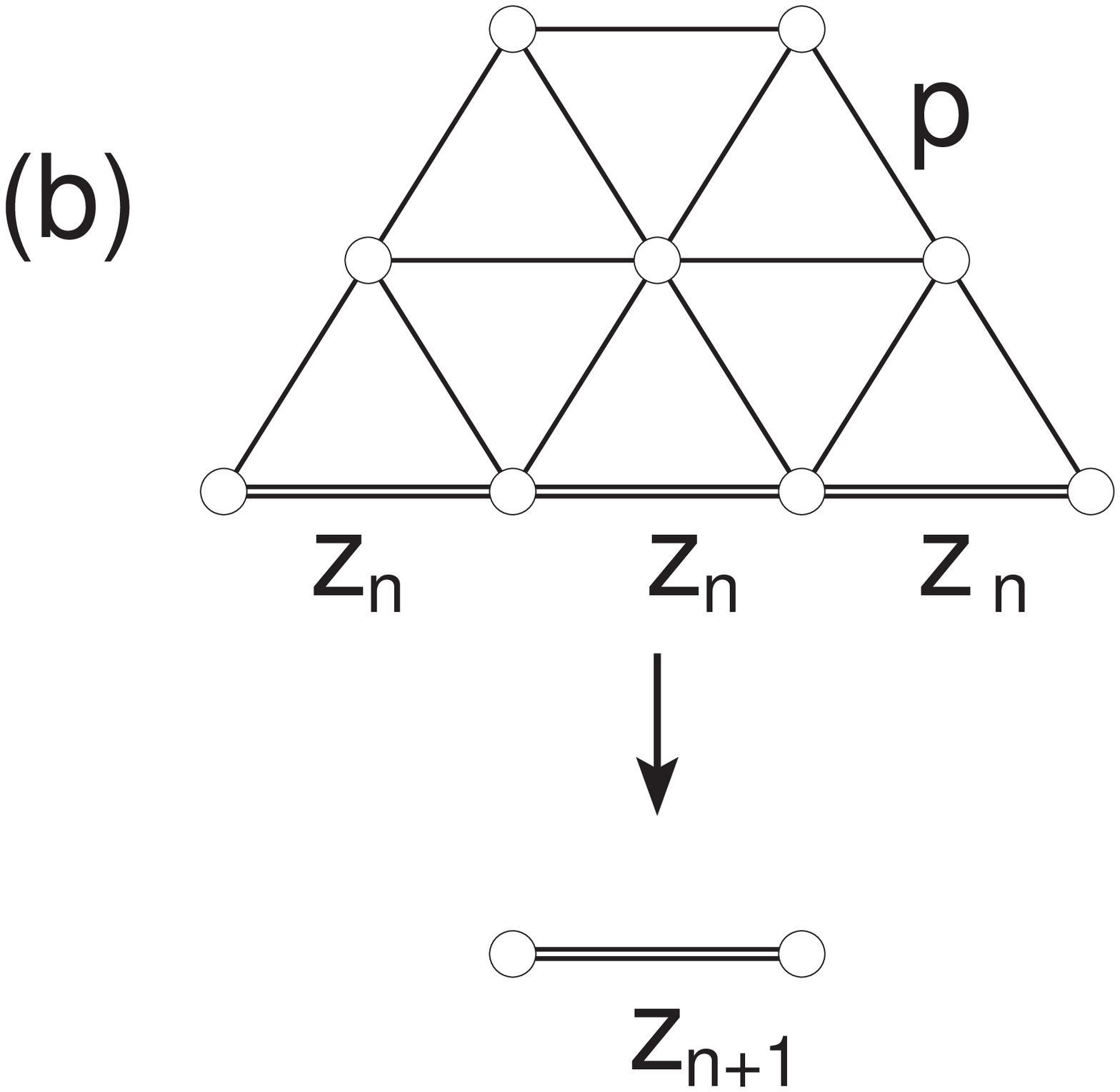}\\
\includegraphics[width=0.28\textwidth]{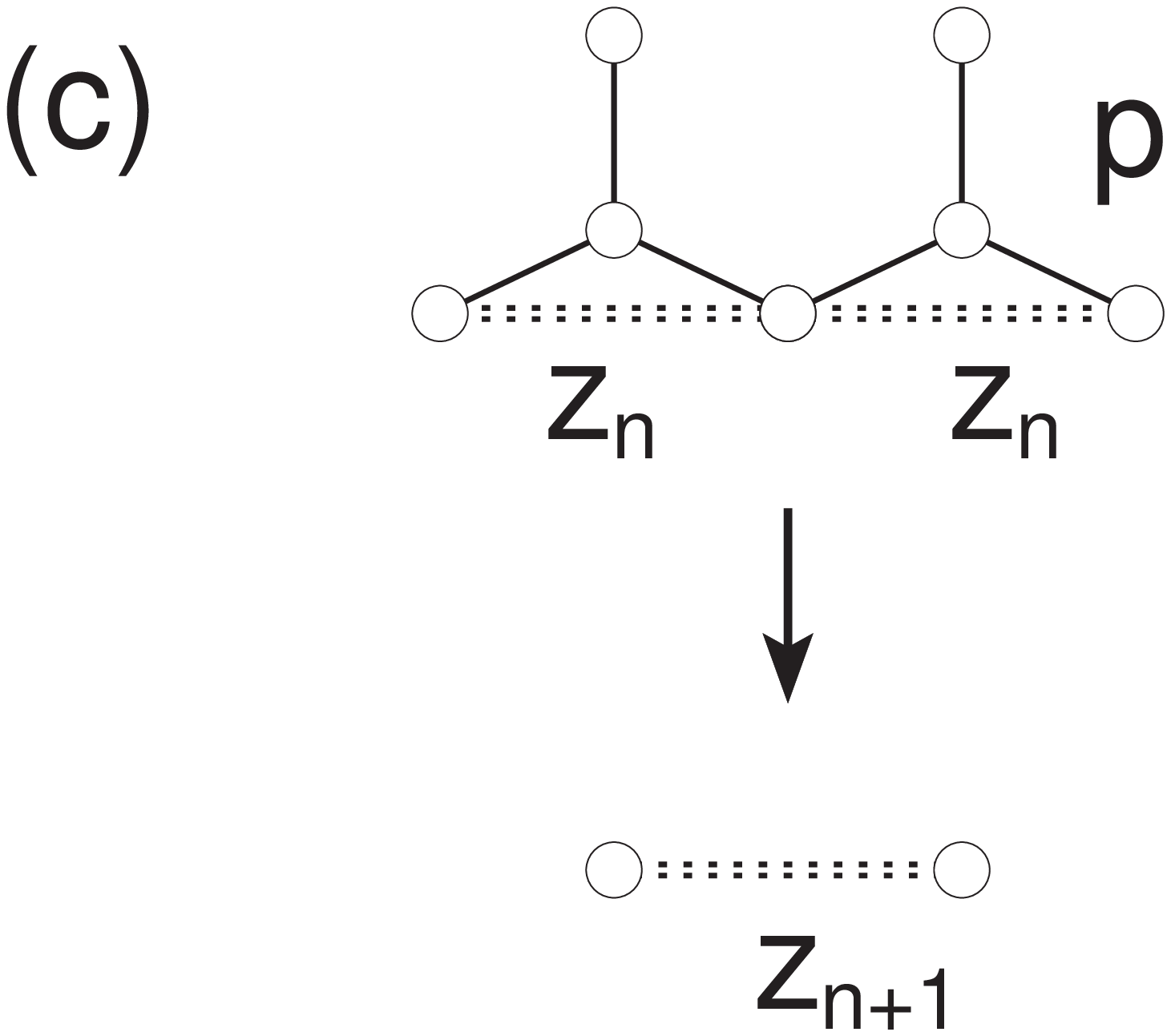}
\includegraphics[width=0.28\textwidth]{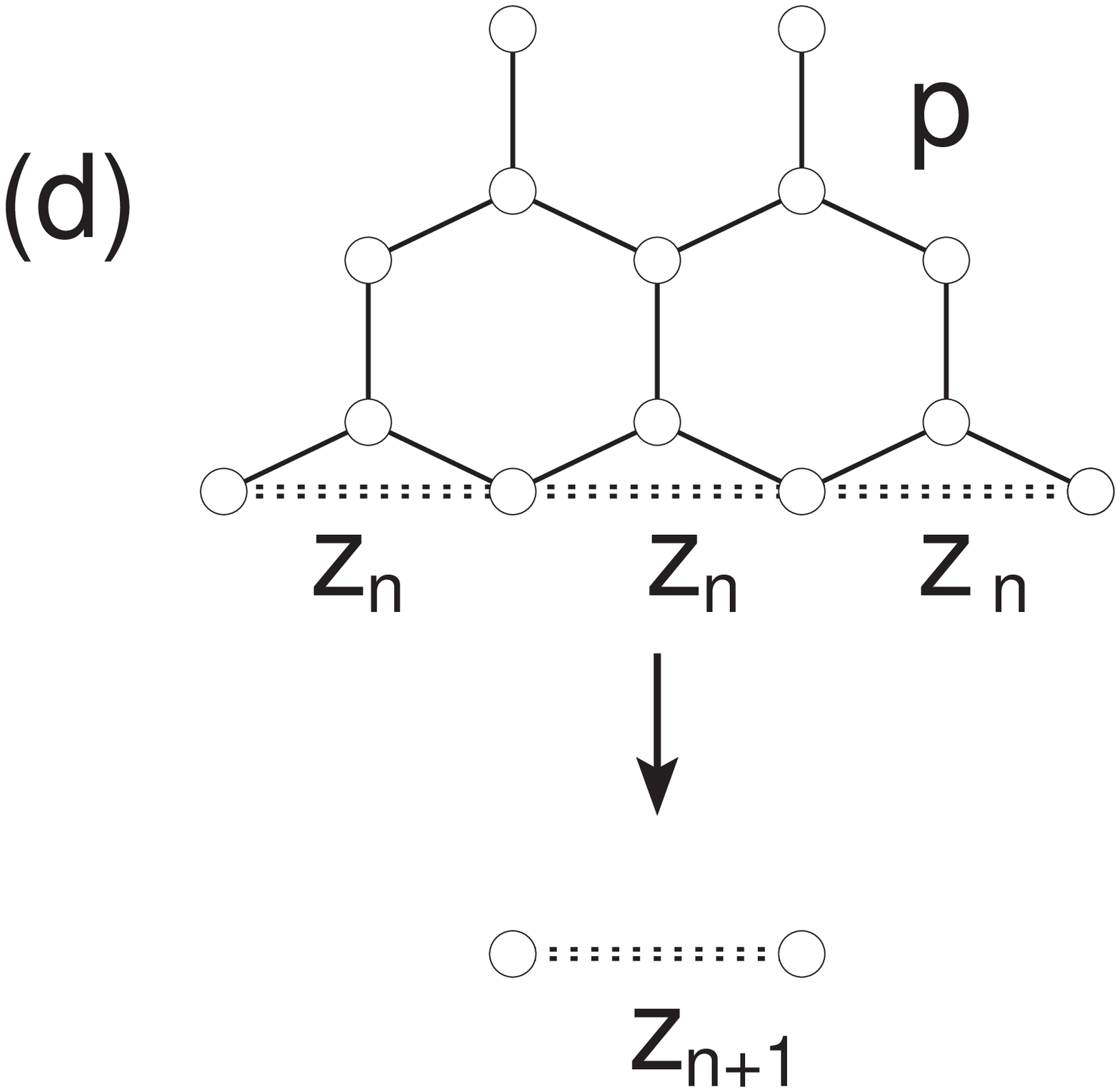}\\
\includegraphics[width=0.28\textwidth]{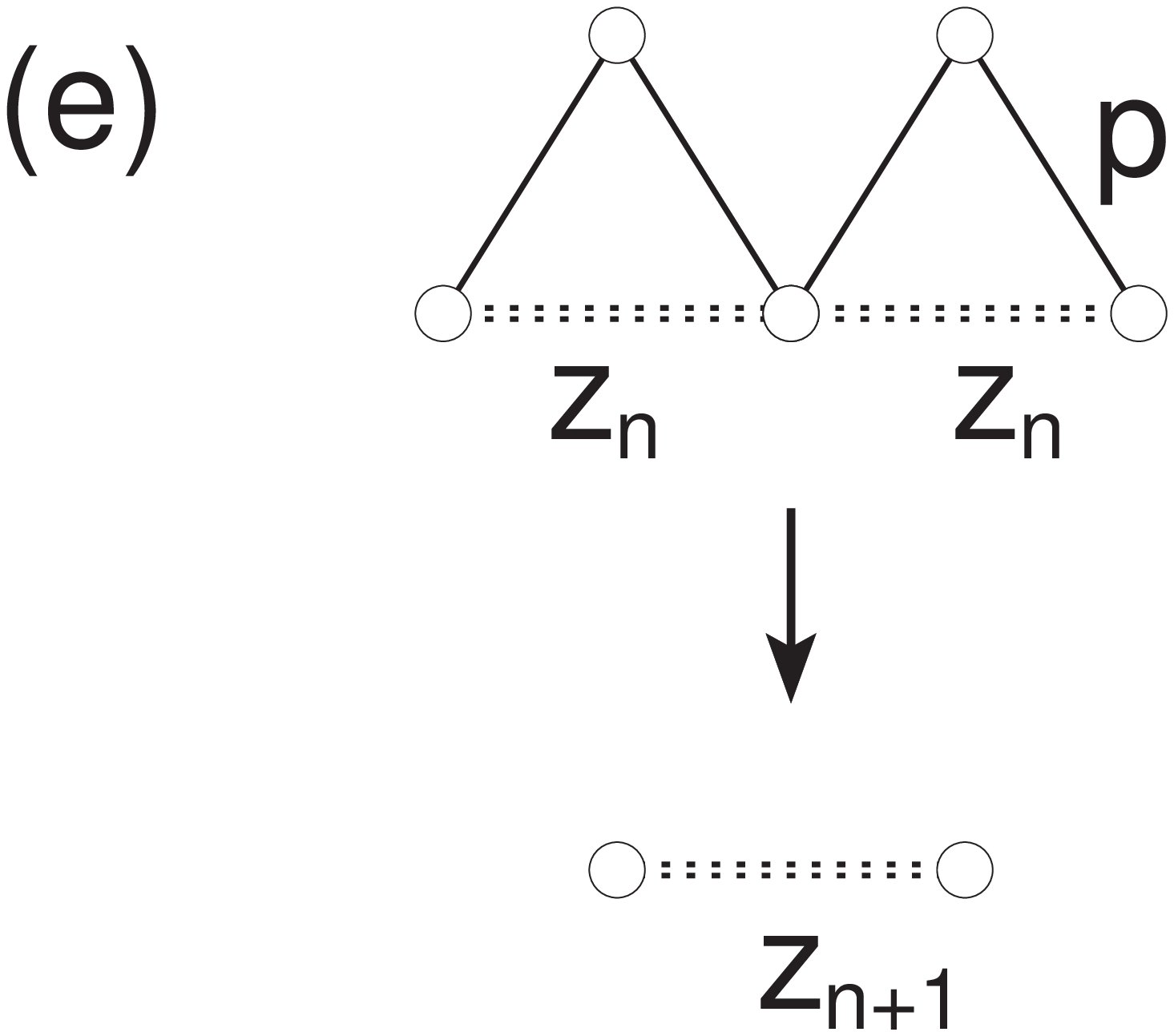}
\includegraphics[width=0.28\textwidth]{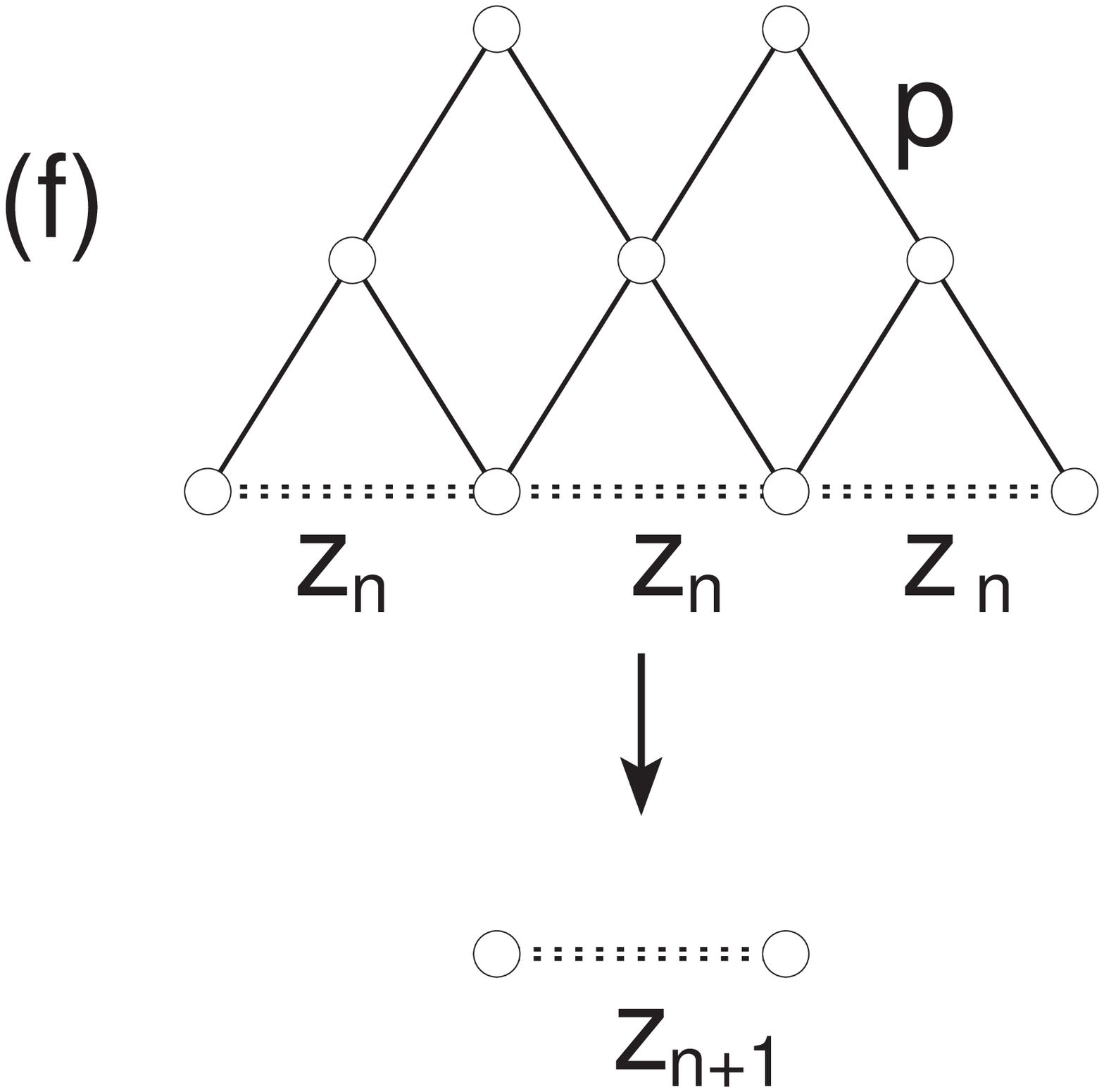}
\end{center}
\caption{(a) A variation of the RG scheme in
\fref{fig:ebt}(b). Here it describes the triangular lattice with width
$w=1$. (b) A larger cell with $w=2$ for the triangular lattice.
The RG scheme for the honeycomb lattice with (c)
$w=1$ and (d) $w=2$ can be constructed in the same way,
as well as that for the square lattice with (e)
$w=1$ and (f) $w=2$.
The double lines represent coarse-grained effective bonds and
the dotted lines in (c) to (f) mean that the connections do not
correspond to any bare interaction.
}
\label{fig:reg}
\end{figure}

Although the above RG scheme is devised to investigate a
hierarchical structure, we show in this work that it can be applied to
non-hierarchical planar lattices as well. In \fref{fig:reg}(a),
we present a variation of the RG scheme shown above.
The similarity is obvious: we have taken away only one bond out of those in
\fref{fig:ebt}(b), and this is meant to describe the triangular lattice.
It leads us to the following recursion,
\begin{equation}
z_{n + 1} = p  +  (1 - p) \left[ p  +  (1 - p)z_n \right]^2.
\label{eq:rec1}
\end{equation}
Again, the bond connection in {\em lower} levels, composed of $p$ and $z_n$,
is converted to a single bond with $z_{n+1}$ at a {\em higher} level. This
distinction of levels might look arbitrary since the bonds in the plane do
not have any hierarchy. However, the important point is that all the
argument above to find a lower bound remains still legitimate from this
viewpoint.
Solving \eref{eq:rec1} for $z_{n + 1} = z_n = z_{\infty}$, we find that
\[ z_{\infty} = \frac{p(1 + p - p^2)}{(1 - p)^3} \]
and consequently, $z_{\infty}=1$ at $p^{\ast} = 1 - 1/\sqrt{2} \approx 0.293$.
Comparing this to the exact bond-percolation threshold in the triangular
lattice, $p_c^{\rm t} \approx 0.347$~\cite{sykes}, we see that our method
indeed yields a lower bound.
We now extend the cells to be renormalized by adding one more level.
That is, let us denote the width of the cell as $w$ and consider the case of
$w=2$. For the triangular lattice, the shape of such a larger cell is
given in \fref{fig:reg}(b). By enumerating all the possible cases, the
recursion relation is obtained as
\begin{eqnarray*}
z_{n+1} &=& 3p^9z_n^3 - 25p^8z_n^3 + 90p^7z_n^3 - 182p^6z_n^3\\
&& + 224p^5z_n^3 - 168 p^4z_n^3 + 70p^3z_n^3 - 10p^2z_n^3\\
&& - 3pz_n^3 + z_n^3 - 7p^9z_n^2 + 53p^8z_n^2\\
&& - 171p^7z_n^2 + 303p^6z_n^2 - 315p^5z_n^2 + 187p^4z_n^2\\
&& - 53p^3z_n^2 + p^2z_n^2 + 2pz_n^2 + 5p^9z_n\\
&& - 33p^8z_n + 89p^7z_n - 121p^6 z_n + 79p^5z_n\\
&& - 11p^4z_n - 13p^3z_n + 5p^2z_n - p^9 + 5p^8\\
&& - 8p^7 + 12 p^5 - 8p^4 - 4p^3 + 4p^2 + p,
\end{eqnarray*}
and we find its limiting value as
\begin{eqnarray*}
z_{\infty} &=&  \frac{ F_1(p) - \sqrt{F_2(p)} }{F_3(p)}.
\end{eqnarray*}
with
$F_1(p) \equiv 4p^6 - 16p^5 + 21 p^4 - 6p^3 - 6p^2 + 2p + 1$,
$F_2(p) \equiv 4p^{12} - 40p^{11} + 176p^{10} - 400p^9 + 653p^8 -
508p^7 + 48p^6 + 236p^5 - 126p^4 - 32p^3 + 20p^2 + 8p + 1$,
and
$F_3(p) \equiv 6p^6 - 32p^5 + 66p^4 - 64p^3 + 26 p^2 - 2$.
The solution of $z_{\infty} = 1$ is found at $p^{\ast} \approx 0.300$, which is
an improved lower bound compared to the previous one, $p = 1-1/\sqrt{2}
\approx 0.293$, even though the convergence turns out to be rather slow.

If we also regard the honeycomb lattice as hierarchical, we can consider an
RG scheme as depicted in \fref{fig:reg}(c). By calculating the probabilty
for any of the leftmost points to connect to any of the rightmost points
within the cell, we find
\begin{equation}
z_{n + 1} = \left[ p  +  (1 - p)z_n \right]^2.
\label{eq:honey}
\end{equation}
By a little algebra as above, we find $z_{\infty} = p^2 / (1 - p)^2$, which
becomes one at $p^{\ast} = 1/2$. Again, this is lower than the exact value
$p_c^{\rm h} \approx 0.653$~\cite{sykes}. One may expect an improved
estimate by considering a larger cell shown in \fref{fig:reg}(d), which
leads to
\begin{eqnarray*}
z_{n+1} &=&-4 p^7 z_n^3 + 17 p^6 z_n^3 - 26 p^5 z_n^3 + 15 p^4 z_n^3\\
&&- p^2 z_n^3 - 2 p z_n^3 + z_n^3\\
&&+ 10 p^7 z_n^2 - 35 p^6 z_n^2 + 40 p^5 z_n^2 - 12 p^4 z_n^2\\
&& - 4 p^3 z_n^2 - p^2 z_n^2 + 2 p z_n^2 - 8 p^7 z_n\\
&& + 21 p^6 z_n - 12 p^5 z_n - 6 p^4 z_n + 4 p^3 z_n\\
&& + p^2 z_n + 2 p^7 - 3 p^6 - 2 p^5 + 3 p^4 + p^2.
\end{eqnarray*}
The limiting solution is
\[ z_{\infty} = \frac{G_1(p)-\sqrt{G_2(p)}}{G_3(p)}, \]
where
$G_1(p) \equiv - 6p^5 + 6p^4 + 4p^3 - p^2 - 2p - 1$,
$G_2(p) \equiv 4p^{10} - 16p^9 + 24p^8 - 12p^7 + 12p^6 - 20p^5 - 11p^4 +
4p^3 + 10p^2 + 4p + 1$,
and $G_3(p) \equiv 8p^5 - 18p^4 + 8p^3 + 4p^2 -2$. We find that $z_{\infty}
= 1$ at $p^{\ast} \approx 0.537$. Using the duality relation $p_c^{\rm t} +
p_c^{\rm h} = 1$~\cite{sykes}, we may turn this result
to an {\em upper} bound of
the bond-percolation threshold for the triangular lattice. That is, our
method gives a possible region of the threshold as $0.300 \le p_c^{\rm t}
\le 0.463$, or equivalently, $0.573 \le p_c^{\rm h} \le 0.700$.

A more interesting case is found by considering the horizontal bonds in
\fref{fig:reg}(a) and \fref{fig:reg}(b) as
fictitious [\fref{fig:reg}(e) and \fref{fig:reg}(f)].
This corresponds to the square lattice,
and the interaction in the horizontal direction will appear
only as an effective one mediated by shorter bonds.
Then we can simplify \eref{eq:rec1} as
\begin{equation*}
z_{n + 1} = \left[ p  +  (1 - p)z_n \right]^2,
\end{equation*}
which happens to be the same as \eref{eq:honey}. Therefore, we find
$p^{\ast} = 1/2$ once again, but this value is identical to the
exact value for the bond-percolation problem in the square
lattice~\cite{kesten}.
Since this method is supposed to give a lower bound, it should not be
possible to improve this result further, so it will be worth checking
whether this value really remains unchanged for a larger cell. From a larger
cell depicted in \fref{fig:reg}(f), we obtain a recursion
\begin{eqnarray*}
z_{n+1} &=& - 3p^6 z_n ^3 + 14p^5 z_n^3 - 25p^4 z_n ^3 + 20p^3 z_n ^3\\
&& - 5p^2 z_n ^3 - 2 p z_n ^3 +  z_n^3 + 7p^6 z_n ^2\\
&& - 28p^5 z_n ^2 + 40p^4 z_n ^2 - 22p^3 z_n ^2 + p^2 z_n^2\\
&& + 2p z_n ^2 - 5p^6 z_n + 16p^5 z_n  - 14p^4 z_n\\
&&  + 3p^2 z_n  + p^6 - 2p^5 - p^4 + 2p^3 + p^2,
\end{eqnarray*}
and find its limiting value as
\[ z_{\infty} = \frac{H_1(p) - \sqrt{H_2(p)}}{H_3(p)} \]
with
$H_1(p) \equiv 4p^4 - 6p^3 - p^2 + 2p + 1$,
$H_2(p) \equiv 4p^8-16p^7+27p^6-12p^5-15p^4+6p^2+4p+1$,
and
$H_3(p) \equiv 6p^4-16p^3+12p^2-2$.
The critical
value making $z_{\infty}=1$ is also $p^{\ast} = 1/2$, as expected.
The fact that $p^{\ast}$ does not change with $w$
could be an evidence that the bond-percolation threshold is located
exactly at $p = 1/2$ for the square lattice.

In addition, we can argue that the connection probability over distance $l$
would be roughly determined by $\left( z_{\infty} \right)^l = e^{ l \log
z_{\infty} }$ near the critical point. In other words, the correlation
length would be written as $\xi \sim -1/\log z_{\infty}$. The slope of
$z_{\infty}$ around $p=p_c$ does not vanish in every case considered above,
so it generally behaves as $z_{\infty} \sim a (p-p_c)+1$
where $a \equiv \left. \partial z_{\infty}/\partial p \right|_{p=p^{\ast}}
\sim O(1)$ at $p = p_c - \epsilon$ with positive $\epsilon \ll 1$.
Therefore, we see that
\begin{eqnarray*}
\xi &\sim& -\frac{1}{\log z_{\infty}} \sim -\frac{1}{\log [a (p-p_c)+1]}\\
&\approx& (p_c - p)^{-1},
\end{eqnarray*}
by using $\log(1-a \epsilon) \approx -a \epsilon$.
Since the correlation length is assumed to diverge as $\xi \sim
|p-p_c|^{-\nu}$, this argument gives us an approximate estimate of the
critical exponent as $\nu \approx 1$, which is an underestimate
compared to the exact value, $\nu = 4/3$~\cite{nijs}.
It is worth noting that this RG scheme does
not make use of any explicit scaling transformation: we do not zoom up or
zoom down the system at criticality as usually found in RG
studies~\cite{nie,rey77,rey80}. In arguing the value of $\nu$, therefore, we
evaluate it directly in units of the given lattice spacing instead of
any zooming ratio. By setting $z_n = z_{n+1}$, in a sense,
it is the translational invariance that we are actually exploiting in this
study.

In summary, we have shown that the RG scheme devised for
a hierarchical structure can be also applied to the 2D lattices
even though they are not hierarchical. It generally yields a lower
bound, but correctly predicts the bond-percolation threshold for
square lattice. We have also approximately estimated $\nu \approx 1$.
This method is more related to the translational invariance rather than
to the scaling invariance at criticality.

\ack
We are grateful for support from the Swedish Research Council with
Grant No. 621-2008-4449.

\section*{References}

\end{document}